\documentclass[12pt,preprint]{aastex}

\usepackage{natbib}
\usepackage{ulem}
\newcommand{\no}[1]{#1}

\slugcomment{Accepted for publication in The Astrophysical Journal}
\shorttitle{ $\gamma$-ray emission from SOL2010-06-12T00:57}

\begin{document}


\title{{\it{Fermi}} Detection of $\gamma$-ray emission from the M2 Soft X-ray Flare on 2010 June 12}


\author{
M.~Ackermann\altaffilmark{2}, 
M.~Ajello\altaffilmark{3}, 
A.~Allafort\altaffilmark{3}, 
W.~B.~Atwood\altaffilmark{4}, 
L.~Baldini\altaffilmark{5}, 
G.~Barbiellini\altaffilmark{6,7}, 
D.~Bastieri\altaffilmark{8,9}, 
K.~Bechtol\altaffilmark{3}, 
R.~Bellazzini\altaffilmark{5}, 
P.~N.~Bhat\altaffilmark{10}
% ...
}
% ...
\altaffiltext{2}{Deutsches Elektronen Synchrotron DESY, D-15738 Zeuthen, Germany}
\altaffiltext{3}{W. W. Hansen Experimental Physics Laboratory, Kavli Institute for Particle Astrophysics and Cosmology, Department of Physics and SLAC National Accelerator Laboratory, Stanford University, Stanford, CA 94305, USA}
\altaffiltext{4}{Santa Cruz Institute for Particle Physics, Department of Physics and Department of Astronomy and Astrophysics, University of California at Santa Cruz, Santa Cruz, CA 95064, USA}
\altaffiltext{5}{Istituto Nazionale di Fisica Nucleare, Sezione di Pisa, I-56127 Pisa, Italy}
\altaffiltext{6}{Istituto Nazionale di Fisica Nucleare, Sezione di Trieste, I-34127 Trieste, Italy}
\altaffiltext{7}{Dipartimento di Fisica, Universit\`a di Trieste, I-34127 Trieste, Italy}
\altaffiltext{8}{Istituto Nazionale di Fisica Nucleare, Sezione di Padova, I-35131 Padova, Italy}
\altaffiltext{9}{Dipartimento di Fisica ``G. Galilei", Universit\`a di Padova, I-35131 Padova, Italy}
\altaffiltext{10}{Center for Space Plasma and Aeronomic Research (CSPAR), University of Alabama in Huntsville, Huntsville, AL 35899, USA}
% ...
\setcounter {footnote}{0}
\setcounter {mpfootnote}{0}
%\setcounter {affil}{0}

%\altaffiltext{1}{University of Alabama in Huntsville, NSSTC, 320 Sparkman Drive, Huntsville, AL 35805, USA}


\begin{abstract}
The {\it{GOES}} M2-class solar flare, SOL2010-06-12T00:57, was modest in many respects yet exhibited remarkable acceleration of energetic particles. The flare produced an $\sim$50 s impulsive burst of hard X- and $\gamma$-ray emission up to at least 400 MeV observed by the {\it{Fermi}} GBM and LAT experiments. {\no The remarkably similar hard X-ray and high-energy $\gamma$-ray time profiles suggest that most of the particles were accelerated to energies $\gtrsim$300 MeV with a delay of $\sim$10 s from mildly relativistic electrons, but some reached these energies in as little as $\sim$3 s.}
The $\gamma$-ray line fluence from this flare was about ten times higher than that typically observed from this modest {\no GOES} class of X-ray flare.
There is no evidence for time-extended $>$100 MeV emission as has been found for other flares with high-energy $\gamma$ rays.

\end{abstract}





\section{Introduction}

The Sun is capable of accelerating electrons and ions to relativistic energies on time scales as short as a few seconds during solar flares.   This conclusion has been reached based on observations of the X rays, microwaves, $\gamma$ rays, and neutrons produced when the flare-accelerated particles interact in the solar atmosphere \citep{forr83,kane86}.  The first reported observation of $\gamma$ rays with energies above 10 MeV was made with the {\it{Solar Maximum Mission}} ({\it{SMM}}) spectrometer during the 1981 June 21 flare \citep{chup82}. Most of the $\gamma$-ray emission occurred within an $\sim$70 s period and was followed minutes later by detection of high-energy neutrons at the spacecraft.  Although it was clear from the neutron timing observations that protons were accelerated to energies in excess of 100 MeV, it was not possible from $\gamma$-ray spectroscopic studies to conclude that the protons reached the energies above $\sim$300 MeV necessary to produce the characteristic spectrum from pion-decay radiation \citep{murp87}.

\citet{forr85,forr86} provided clear spectroscopic evidence for pion-decay emission during both the prompt and delayed emission phases of the 1982 June 3 flare. This emission in the second phase was confirmed by the presence of the 0.511-MeV annihilation line from the decay of positively charged pions that had a time profile similar to the high-energy $\gamma$-rays \citep{shar83}.

Since the era of those early measurements, improved spectrometers have detected the presence of pion-decay emission in several more flares and in two cases have observed $\gamma$-ray emission up to at least 1 GeV \citep{kanb93,akim96,vilm03,kuzn11}. \citet{ryan00} and \cite{chup09} have reviewed these and other examples of high-energy $\gamma$-ray emission in flares.  Most of these high-energy emissions have been observed over tens of minutes to hours leading to the designation of a class known as `Long Duration Gamma-Ray Flares'.  Several processes have been suggested to explain such particle acceleration to high energies in the solar environment \citep{ell85,petr94,park97,ryan00,asch04,chup09}.

Typically such high-energy $\gamma$-ray emission has been associated with intense soft X-ray flares recorded by {\it{GOES}} having peak powers exceeding $10^{-4}$ W m$^{-2}$ (X class).   It is of interest to determine whether less energetic solar flares have the capability of accelerating electrons to energies $\gtrsim$100 MeV and protons to energies  $\geq$300 MeV.   Out of 24 flares observed at $\geq10$ MeV energies by the $\gamma$-ray spectrometer on {\it{SMM}} only 3 had {\it{GOES}} classifications of M5 or less \citep{vest99}.  Similarly, most flares emitting $\gamma$-ray lines have been associated with intense soft X-ray emission: out of 65 $\gamma$-ray line flares observed by {\it{SMM}} only three have been associated with flares having a {\it{GOES}} classification of M5 or less.  Before detection of this flare, the {\it{Ramaty High Energy Solar Spectroscopic Imager}} ({\it{RHESSI}}) observed only one nuclear-line flare out of 20  below the M5 class \citep{shih09} and that was characterized as C9.7.  The smallest {\it{GOES}} class flare for which detection of nuclear $\gamma$-rays has been claimed was the C7 flare observed by the COMPTEL  Compton telescope on the {\it{Compton Gamma-Ray Observatory}} ({\it{CGRO}}) \citep{young01}.

With the launch of the {\it Fermi} mission in 2008, it is now possible to make the high-sensitivity measurements necessary to detect $\geq30$ MeV $\gamma$ rays in the weakest flares. {\it Fermi} is comprised of two instruments: the Gamma-ray Burst Monitor (GBM) \citep{meeg09} sensitive from $\sim 8$ keV up to 40 MeV and covering the energy band of nuclear $\gamma$-ray line emission; and the Large Area Telescope (LAT) \citep{atwo09} operating from 20 MeV to more than 300 GeV and covering most of the pion-decay emission energy range.  In its normal sky survey mode the LAT observes the Sun for only $\sim35$ min every 3 hours.  It was therefore fortunate to observe the M2 {\it{GOES}}-class flare on 2010 June 12 (SOL2010-06-12T00:57).  This flare produced nuclear line emission about an order of magnitude higher than typical for this {\no magnitude} soft X-ray flare.  Although the flare lasted only about 2 min, it appears to have accelerated electrons and/or protons to energies $\geq$300 MeV based on detection of $\gamma$ rays with energies $>$100 MeV.  One of the {\no key} features of this observation is that {\no the} high-energy emission is delayed by {\no $\sim 5-10$ s} relative to the {\no 500 -- 1000} keV bremsstrahlung.  This suggests that the acceleration of hundreds of keV electrons and hundreds of MeV electrons and/or protons {in the chromosphere took place {\no within 10 s of each other}.
A weak solar energetic particle (SEP) event observed by {\it{GOES}} followed this flare.

In the next two sections we discuss the capabilities of the GBM and LAT for detecting $\gamma$-rays from solar flares.  These are followed by $\S$4 describing the observations including spectroscopic studies in the nuclear energy range with the GBM and in the high-energy domain with the LAT.  As most high-energy $\gamma$-ray flares previously observed exhibited an extended emission phase, we then discuss a search for such emission in this flare in the next section.  In the final section, $\S$5, we summarize our conclusions and discuss the results.

\section{GBM Capability for Flare Observations}

The Gamma-ray Burst Monitor (GBM) was designed to observe gamma-ray bursts but has useful capabilities for other sources such as SGRs (soft $\gamma$-ray repeaters), pulsars, X-ray binaries using the Earth occultation technique, Terrestrial Gamma-ray Flashes (TGFs), and solar flares \citep{meeg09}.  The GBM is comprised of twelve sodium iodide (NaI) detectors measuring the energy range from $\sim$8 keV to 1 MeV, and two bismuth germanate (BGO) detectors covering the range from $\sim$200 keV to 40 MeV.   The detectors are arranged to collectively view the entire sky not blocked by the Earth.

In this paper we use the BGO detector viewing the Sun to observe the nuclear line and continuum emission from solar flare SOL2010-06-12T00:57. Each BGO detector is a cylinder of length and diameter 12.7 cm, viewed by a photomultiplier tube at each end.  Its effective area for detecting photons ranges from 160 cm$^2$ to 200 cm$^2$, depending on energy and direction of incidence.   Importantly for measuring solar flare
nuclear lines, the high Z of bismuth, the high density of BGO, and the large volume of the GBM BGO detectors result in a high probability for absorbing the full incident photon energy: $\sim$ 67\% at 1 MeV, 50\% at 3 MeV and 40\% at 10 MeV.

Several types of data are produced by the GBM.  There are two temporally-binned types, Continuous Time  (CTIME) and Continuous Spectroscopy (CSPEC), as well as Time-Tagged Events (TTE).    For intense solar flares, TTE data are likely to be lost due to a bandwidth limit between the GBM and the spacecraft solid-state recorders.   Therefore for solar-flare spectral analysis the appropriate data type is CSPEC.    CSPEC has 128 quasi-logarithmic energy bins and 4.096 s temporal resolution, which is improved to 1.024 s when the flight software triggers on a statistically significant rate increase.

The detector performance was measured before launch using X-rays from the BESSY synchrotron radiation facility, radioactive isotopes including the $^{12}$C line at 4.43 MeV (from an $^{241}$Am/$^9$Be source), and $\gamma$-rays up to
17.6 MeV produced via (p,$\gamma$) reactions using a Van de Graaff accelerator.   These data have been used to calibrate the gain, resolution, and response of the detectors \citep{biss09, meeg09}. 
{\no The automatic gain software measures and adjusts the gains of the BGO detectors using the 2.223 MeV line present in the background from capture of cosmic-ray produced neutrons on H in the spacecraft's propellant. In order to improve the gain solution for analysis of the June 12 flare data we measured the centroid of the solar neutron-capture line.  This resulted in a 1\% gain change that is small relative to the $\sim$ 8\% resolution of the GBM at the 2.223 MeV line.}
The best-fit energy of the positron annihilation line (0.511 MeV) from the flare was found to be 0.530 $\pm$ 0.007 MeV, in disagreement with fits of the instrumental background line. However, both lines are superimposed on strong continua making it difficult to determine the lines' peak position.   Further investigation of the BGO calibration below 1 MeV is planned, particularly since TGF spectra also show a shifted positron annihilation line \citep{brig11}.

Spectral fitting is performed using the forward-folding technique with an assumed  parametrized photon spectrum folded through a detector response matrix (DRM) to produce a counts spectrum; the parameters are adjusted to obtain the best fit to the observed counts spectrum.  The DRM is based on Geant4 simulations of the GBM detectors and mass model of the satellite and has been validated by comparison with the calibration data \citep{hoov08}.

\section{LAT Capability for Flare Observations}

The Large Area Telescope (LAT) is a pair-conversion telescope designed to detect gamma rays from 20 MeV up to more than 300 GeV \citep{atwo09}. It is made up of a 4 $\times$ 4 matrix of identical towers, each one comprised of a tracker with layers of Silicon Strip Detectors (SSD) alternating with foils of high-Z converter (tungsten), and a calorimeter with logs of CsI arranged in a `hodoscopic' configuration so that the energy deposition is imaged in three dimensions. The array of towers is surrounded by an Anti-Coincidence Detector (ACD) made up of 89 plastic scintillator tiles with a 5 $\times$ 5 array on the top and 16 tiles on each of the 4 sides.  {\no To reduce the impact of the self-veto events caused by calorimeter back-splash, each ACD tile is typically $<$1000~cm$^2$, depending on its position in the array.} A $\gamma$-ray passes through the ACD with small probability of interaction and can convert into an electron-positron pair that is tracked in the SSD.  The energy of a photon below about 100 MeV can be estimated by multiple-Coulomb scattering of the electrons in the tungsten converters and SSD.   Energies of higher-energy photons are measured by `total' absorption or modeling the shower profile in the calorimeter.   
In the standard {\it{Fermi}} sky-survey mode {\no the spacecraft rocks  50$^{\circ}$ north and south in celestial declination from the zenith so that} each region of the sky is viewed for $\sim$30-40 min every two orbits; therefore the LAT's large aperture (2.4 sr) and effective area provides the capability to sensitively monitor solar activity with a duty cycle of 15-20\%.

The ACD is used to reject the large background of charged cosmic-radiation and secondaries from the spacecraft and Earth's atmosphere.  The threshold for this veto is nominally set at 45\% of the amplitude of a minimizing ionizing singly-charged particle traversing a tile in the ACD (i.e. set to $\sim$800 keV).  If this threshold is exceeded and if the ACD tile hit was adjacent to a tracker tower that caused the event to trigger, the event will be vetoed unless a sufficiently high energy is deposited in the calorimeter ($>$100 MeV in one or more crystals).  This veto inhibition ensures that backsplash does not cause very high-energy $\gamma$ rays to self-veto.  Of the vetoed events, 2\% are telemetered to the ground for diagnostic purposes.

Solar flares can {\no emit} intense flux{\no es} of tens of keV X rays. {\no About 20\% of the X rays at these energies} can penetrate the thermal blanket and micrometeoroid shield and reach the ACD, {\no depositing part of their energy in the illuminated tile.}

Several of these photons can arrive within the 0.4 $\mu$s anti-coincidence veto shaping time
(pulse pile-up) to yield a high total-energy loss.  It is possible in the largest of flares that this energy loss could exceed the 800 keV veto threshold and information about a valid photon event would not be transmitted to the ground.  None of the solar flares detected to date have been sufficiently intense to cause this data loss to occur.

Pulse pile-up of hard X rays from some flares has, however, affected the ground analysis of LAT data, in the classification of events as $\gamma$ rays or background.  ACD tile hits are registered (i.e. included in an event) whenever there is $>$100 keV energy deposition integrated over the $\sim$3 $\mu$s peaking time of the front-end amplifier (as opposed to the 800 keV threshold for an ACD veto).  Pulse pile-up during even modest flares generates such tile hits.  In order to achieve the highest sensitivity for studying celestial sources, the current LAT ground-processing software rejects $\gamma$-ray events with a high ratio between the number of ACD tile hits (i.e., the number of tiles with energy deposition $>$100 keV) and the energy measured by the calorimeter.  Thus, $\gamma$ rays arriving during relatively intense flares have a high likelihood of being rejected from the standard LAT data products.  This is the case for the 2010 June 12 flare as we will discuss in $\S$4.2 below.

As a consequence of this feature, the LAT Low-Energy (LLE) technique \citep{pela10} was adapted for analyzing data during this flare.  LLE event selection uses less discriminating criteria than the standard ground-processing and is not affected by ACD tile hits $>$100 keV.  The primary requirement is that the candidate $\gamma$-ray event have at least one track and a reconstructed energy larger than 30 MeV. {\no Due to these less discriminating criteria the off-axis angle for LLE events can be as large as $\sim$80$^{\circ}$ compared with $\sim$68$^{\circ}$ in the standard LAT data products.} In addition, only $\gamma$-rays whose estimated arrival directions were within 20$^{\circ}$ of the Sun are included for analysis. The 20$^{\circ}$ angular restriction was chosen based on Monte Carlo simulations of the LAT point-spread function at these energies.  A time series of events is then constructed from which background intervals on either side of the flare are defined and a linear or quadratic interpolation is used to estimate the background during the flare.  A Detector Response Matrix (DRM) for the solar location during the flare is created using a custom Monte Carlo simulation.  By passing candidate photon models through the DRM, we then fit the background-subtracted data using a version of {\em rmfit 3.4}\footnote[1]{R.S. Mallozzi, R.D. Preece, \& M.S. Briggs, ``RMFIT, A Lightcurve and Spectral Analysis Tool'', Robert D. Preece, University of Alabama in Huntsville, (2008): http://fermi.gsfc.nasa.gov/ssc/data/analysis/user/}, customized for the specific solar flare, and the {\em OSPEX}\footnote[2]{SolarSoft: http://www.lmsal.com/solarsoft/} analysis packages.

Using a Monte Carlo analysis, we estimate that the energy resolution for a source at $\sim$75$^{\circ}$ off axis in our LLE analyses is about 40\%.  We also estimate $\sim$15\% and $\sim$10\% systematic uncertainties in the LLE fluxes $>$30 and 100 MeV, respectively, based on studies of the Vela pulsar. The uncertainty $>$100 MeV is consistent with that found using the standard {\it{Fermi}} analysis tools applied to the pulsar~\citep{vela2}.

\section{June 12 Flare Observations}

The {\it{GOES}} M2.0 class X-ray flare commenced with some low-level activity on 2010 June 12 at 00:30 UT.  Although both {\it{RHESSI}} and {\it{Fermi}} observed the flare, here we only discuss the {\it{Fermi}} observations because {\it{RHESSI}} was not pointed at the Sun during the flare. The flare occurred with {\it{Fermi}} in sunlight and during a relatively low-background portion of its orbit. As seen in Figure \ref{lat_nai}a the $10 - 26$ keV emission recorded by the GBM NaI detectors commenced around 00:55 UT and rose precipitously about 40 s later{\no ; for comparison we also plot the GOES 0.5 -- 4 $\AA$ profile and note that this emission is dominated by 3 keV thermal photons as is reflected in its slower rise and extended tail.} The GBM burst algorithm triggered on the flare at 00:55:05.64 UT and put the instrument in a high-data rate mode for the next 10 min.  The $100 - 300$ keV time profile observed by the GBM's solar facing NaI detector is also plotted in Figure \ref{lat_nai}a. It is clear that the emission peaks more sharply and ends sooner at higher X-ray energies.  Most of the emission observed above 100 keV occurred within an $\sim$50 s interval. The event as viewed by {\it{GOES}} {\no 1 -- 8 $\AA$ channel} ended about 01:30 UT.  The flare originated from active region (AR) 11081 at approximately N23$^{\circ}$W43$^{\circ}$.   White light emission observed by the Helioseismic and Magnetic Imager (HMI) on the {\it{Solar Dynamics Observatory}} ({\it{SDO}}) \citep{oliv11} in a single 45 s exposure, {\no consistent in time with} the hard X-ray emission, reveals two compact footpoints about 10$^4$ km apart. 

\begin{figure}
%\epsscale{.60}
%\plotone{LC.eps}
\plotone{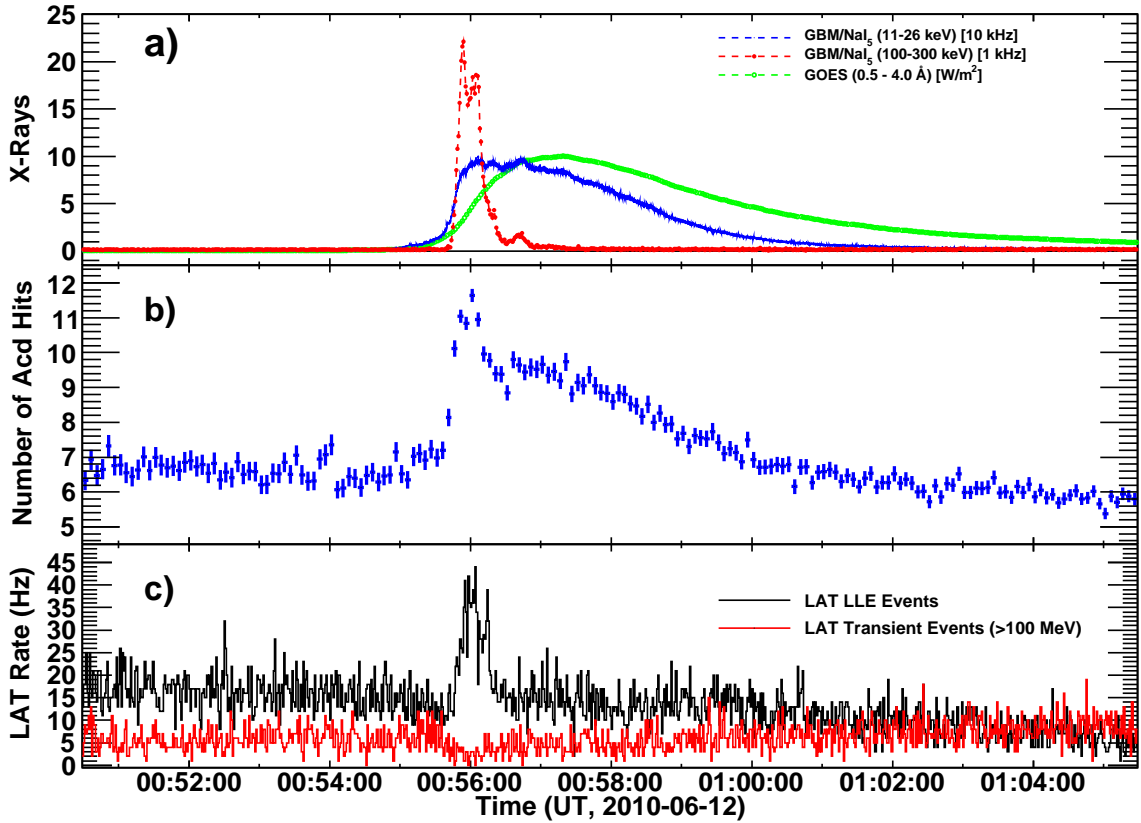}
\caption{Time histories related to the 2010 June 12 solar flare.  a) {\it{GOES}} 0.5 -- 4 $\AA$ rates, and GBM NaI 11 -- 26 keV and 100 -- 300 keV relative rates; b) LAT ACD hit rate $>$100 keV containing contributions from background, $>$100 keV solar flare X rays (impulsive peak) and pulse pile up from 10's of keV solar X rays following the NaI 11 -- 26 keV profile in 1a); and c) LLE and LAT Transient Class event rates. }
\label{lat_nai}
\end{figure}

There was no evidence for significant dead time and/or pulse pile-up effects in the GBM BGO detector facing in the solar direction.  Photons with energies up to $\sim$8 MeV were detected by the GBM during the 50 s peak. 
At the time of the flare {\no the spacecraft was rocking 50$^{\circ}$ to the south so that} the Sun was 76$^{\circ}$ off axis, close to the edge of the field-of-view (FOV) for LLE studies, and the Earth's horizon was entering the FOV. The accompanying hard X-ray emission from the flare was detected in the LAT's ACD. In Figure \ref{lat_nai}b we plot the average number of ACD tile hits as a function of time.  As discussed in the previous section, pulse pile-up from tens of keV hard X-rays exceeded the 100 keV ACD hit threshold.  This is reflected in the broad peak with a maximum near 00:57 UT that has a shape similar to the 11--26 keV emission observed by the GBM NaI detector.  The impulsive peak in the ACD rate is also similar to that observed between 100 and 300 keV by the GBM NaI detector.  There is no evidence for an increase in the number of ACD vetoed events in orbit during the flare, indicating that pulse-pileup from hard X-rays did not exceed the 800 keV veto threshold.  Thus, the overall valid event rate transmitted to the ground for processing was not affected by the ACD response. However, as shown by the red curve in Figure \ref{lat_nai}c there is no evidence for the flare in the well-screened standard LAT data products~\citep[shown in the figure are the events belonging to the ``transient'' event class,][]{atwo09}.  If anything, we see a deficit of events in the standard analysis light curve which is a consequence of the high ACD hit rate $>$100 keV, shown in Figure \ref{lat_nai}b, that caused a significant increase in the number of events that failed the standard quality cut.
%This resulted from the high ACD hit rate $>$100 keV, shown in Figure \ref{lat_nai}b, that caused a significant increase in the number of events that failed the standard quality cut.

It is important to establish convincingly that any high-energy emission observed by the LAT originated at the flare site and was not due to artifacts from the high rates encountered by the instrument.   The most compelling evidence for the solar flare origin is the map of events relative to the position of the Sun.  Plotted in Figure \ref{map}a is the distribution of LLE-selected events with energies $>$30 MeV accumulated 30 s before and after the flare. 
The distribution is affected by the instrument field of view and by the removal of events near the Earth's horizon.
We reduce the contamination from the bright Earth limb selecting only the events with a reconstructed zenith angle less than 100$^\circ$, but, due to the large point spread function at low energy, some residual contamination near the horizon may be due to this limb brightening.

\begin{figure}
\epsscale{.40}
\plotone{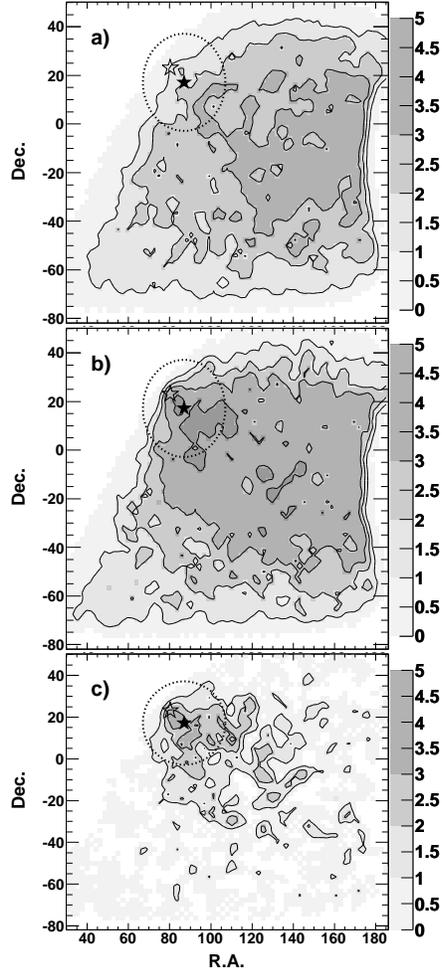}
\caption{Angular distribution of $>$30 MeV $\gamma$ rays relative to the position of the Sun detected using the LLE analysis.
The open star shows the location of the Sun which is close to the is close to the edge of the LAT field of view.
The filled star shows the shifted position due to the fisheye effect discussed in the text. The dotted curve depicts the 20$^{\circ}$ region containing the events used in the analysis.  a) Average of background distribution taken 30 s before and after the flare.  b)  Angular distribution of events observed between 00:55:40 and 00:56:30 UT.  c) Difference between angular distribution observed during the flare and the average background distribution. 
The distributions have been smoothed (with a Gaussian kernel) to reduce statistical fluctuations.}
\label{map}
\end{figure}

In Figure \ref{map}b we plot the distribution accumulated from 00:55:40 to 00:56:30 UT during the flare.   We note that the excess near the Sun's position is biased due to its location near the edge of the LAT's field of view.
This is known as the ``fisheye'' effect, i.e., the tendency of events to be reconstructed with directions closer to the z axis than they should~\citep{Thompson1993}. This bias is relevant only for off-axis events, and it is particularly evident at low energies. The position of the centroid of the counts is shown by the filled star (at R.A.=84$^{\circ}$.39, Dec.=18$^{\circ}$.79, J2000).
It is important to note, that in the routine LAT analyses the contribution to the overall exposure and photon counts from far off-axis events is negligible. In addition to this, owing to the scanning mode of observations with the LAT, persistent sources will have the great majority of their exposure at much smaller inclination angles.

%This effect is well-understood vignetting for sources far off the LAT boresight and is caused by the rapidly falling effective area with increasingly large off-axis angle. Thus, sources at large viewing angles tend to appear closer to the boresight because of this large change in effective area. The position of the centroid of the counts is shown by the filled star (at R.A.=84$^{\circ}$.39, Dec.=18$^{\circ}$.79, J2000).

In Figure \ref{map}c we plot the background-subtracted distribution observed during the flare. There appears to be significant $\gamma$-ray emission consistent with the reconstructed position of the Sun up to energies of $\sim$400 MeV.  We have confirmed the observed 6$^{\circ}$ shift in the position of the centroid from the solar position using a Monte Carlo study.  Using the shift and identifying a $<$ 20$^{\circ}$ acceptance angle (dashed curve in the figure) for LLE events $>$30 MeV maximizes the signal/background. Events that meet these criteria make up the LLE rate (black curve) plotted in Figure~\ref{lat_nai}c. The overall rate decreased, especially after 01:00 UT because the defined aperture approached the exclusion region near the Earth's limb and many events were consequently rejected. The $>$30 MeV LLE time profile during the flare is similar to the 100--300 keV NaI time history.

\subsection{GBM Spectroscopic Studies}
{\no In order to obtain background-subtracted spectra we used GBM/BGO spectra accumulated just before the flare and after 4-min following the flare for background in order to avoid times when there was significant 15 -- 50 keV hard X-ray emission.}  The background-subtracted GBM counts spectrum accumulated over a 50-sec period (00:55:40-00:56:30 UT) during the impulsive phase of the flare is shown in Figure \ref{gbm}.  Although the flare was relatively weak and the BGO detector has only moderate energy resolution, line features are clearly evident in the spectrum.

\begin{figure}
\epsscale{.60}
\plotone{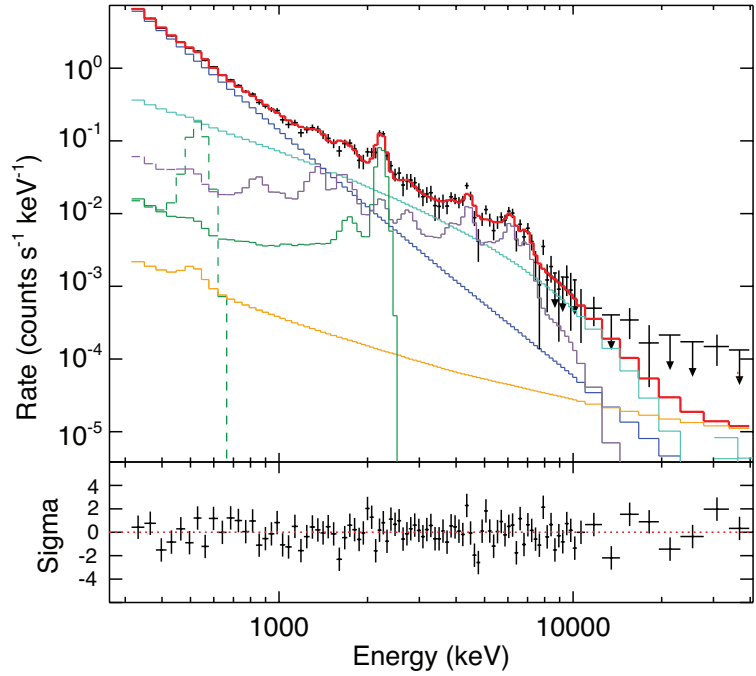}
\caption{Background-subtracted BGO count spectrum accumulated between 00:55:40
and 00:56:30 UT.
The spectrum has been fit by a simple power law (blue histogram) and a flatter
power-law with exponential cut off ({\no light} green histogram) to model
bremsstrahlung by electrons (see text), and with a nuclear de-excitation component (purple) plus 0.511 and 2.223 MeV lines ({\no dashed and dark} green), and a pion-decay (orange) component.}
\label{gbm}
\end{figure}

The spectrum has been fit with a photon model consisting of several components.  We find that the electron-bremsstrahlung component has the shape of power law at low energies that hardens above several hundred keV and then rolls over in the MeV range.  In this paper we fit the bremsstrahlung spectrum with the sum of a low-energy power law and a flat power law with exponential cutoff in the MeV range (blue and {\no light} green curves, respectively, in Figure \ref{gbm}).  Such a complex shape has been observed before in the spectra of several flares detected by {\it SMM} and {\it RHESSI} (Share \& Murphy, in preparation). \citet{mcti90} have found that the magnitude of the hardening above a few hundred keV observed in some flares is larger than that expected for an electron spectrum following a single power law.  \cite{rieg91} describe flattening in the MeV range followed by roll overs above several tens of MeV in the spectra of some flares.  Such features cannot be explained by transport effects alone \citep{petr94} and must be produced by the acceleration mechanism.  \citet{park97} show that these features can be explained by models based on stochastic acceleration by turbulence once loss mechanisms are properly included. Whether such models can explain the 2010 June 12 bremsstrahlung spectrum requires more study.

The nuclear de-excitation lines and continua are represented by a template based on a detailed study of nuclear gamma-ray production from accelerated-particle interactions with elements found in the solar atmosphere \citep{murp09}. 
% {{\no (in Figure \ref{gbm} is rapresented by a purple histogram)}}.  
Such templates depend on the assumed ambient composition and accelerated-particle composition and spectrum. The BGO spectral data for this flare are inadequate to distinguish among templates derived for different ambient abundances, particle spectra and angular distributions. For this reason we used a arbitrary templates based on earlier studies (Share \& Murphy, in preparation). The accelerated particles were assumed to interact in a thick target with a coronal composition \citep{ream95}, but with $^4$He/H = 0.1. We do not take into account transport effects but instead assume that the accelerated particles have a power-law differential spectrum ($dN/dE \propto E^{\beta}$ with $\beta = -4$), coronal elemental abundance (but with an accelerated $\alpha$/p ratio of 0.2), and an angular distribution that is isotropic in the downward hemisphere. {\no As we mentioned above our results are not sensitive to the assumed compositions.}
The photon model also includes Gaussians representing the 0.511 and 2.223 MeV positron-electron annihilation and neutron-capture lines, respectively, and
a pion-decay spectral component.

We present the best-fitting spectral parameters in Table 1, along with estimates of their 1$\sigma$ uncertainties.  We list fluences obtained by integrating over the 50-s time period, except for the 0.511 and 2.223 MeV lines.  The 0.511 MeV line originates from radioactive decays \citep{koz87,koz04} with half lives that can extend up to hours, in addition to a prompt component from positron annihilation following decay of positively charged pions.  The neutron-capture line is also delayed because of the time required for the neutrons to slow down in the solar atmosphere and photosphere and be captured on H, forming deuterium with the release of 2.223 MeV of binding energy from the mass excess.  For these two lines we have fit the background-subtracted spectrum integrated over a total of 250~s.

\begin{deluxetable}{cc}
\tabletypesize{\scriptsize}
\tablecaption{Best-fitting spectral parameters}
\tablewidth{0pt}
\tablehead{
\colhead{Parameter} & \colhead{Value} }
\startdata
PL1 fluence at 300 keV  & 2.85 $\pm$ 0.1 $\gamma$ cm$^{-2}$ keV$^{-1}     $ \\
PL1 index  & 3.31  $\pm$ 0.09 \\
PL2 fluence at 300 keV  & 0.08 $\pm$ 0.02 $\gamma$ cm$^{-2}$ keV$^{-1}$  \\
PL2 index  & $\lesssim$ 1.2 \\
PL2 Exponential Energy & 2400  $\pm$ 800 keV \\
0.511 MeV line fluence\tablenotemark{a}  & 11.3 $\pm$ 2.5 $\gamma$ cm$^{-2}$  \\
2.223 MeV line fluence\tablenotemark{a}  & 21.3 $\pm$ 2.0 $\gamma$ cm$^{-2}$ \\
Nuclear line fluence  & 23.5 $\pm$ 2.5 $\gamma$ cm$^{-2}$ \\
Pion-decay fluence (GBM) $>$ 200 keV & 1.5 $\pm$ 2.5 $\gamma$ cm$^{-2}$ \\
Pion-decay fluence (LAT\tablenotemark{b} ) $>$ 200 keV & 0.62 $\pm$ 0.07 $\gamma$ cm$^{-2}$ \\
Pion-decay fluence (LAT) $>$ 100 MeV & 0.13 $\pm$ 0.015 $\gamma$ cm$^{-2}$ \\
PL3 fluence at 30 MeV  & (9.2 $\pm$ 2.0) $\times 10^{-6} \gamma$ cm$^{-2}$ keV$^{-1} $ \\
PL3 index  & 1.9 $\pm$ 0.2 \\
\enddata
\tablenotetext{a}{Integrated from 00:55:40 to 00:59:50 UT}
\tablenotetext{b}{Computed by extrapolating to low energies the model that best fits LAT data}

%% Text for table notes should follow after the \enddata but before
%% the \end{deluxetable}. Make sure there is at least one \tablenotemark
%% in the table for each \tablenotetext.

\end{deluxetable}

The low-energy power law (PL1) is well defined by the fit. The higher-energy power law (PL2) with exponential rollover is not as well defined because it competes with the nuclear de-excitation line spectrum.  However our studies indicate that this component is required to provide an acceptable fit to the June 12 flare spectrum.  With this component present, the fit is acceptable at the 10\% confidence level based on the $\chi^2$ statistic; without it, the fit is rejected (0.001 \% confidence level).  There is no evidence for pion-decay emission in the BGO spectrum.  As can be seen in Table 1, assuming an hadronic origin,  the LAT-measured pion-decay fluence that we discuss in the next section is about a factor of ten lower than detectable by the GBM BGO detector.

\subsection{Combined LAT and GBM Spectroscopic Studies}

We have obtained a background-corrected LAT count spectrum $>$30 MeV
accumulated during the 50-s period 00:55:40-00:56:30 UT using the LLE data plotted in Figure \ref{lat_nai}c.
%(Figure \ref{stack}). 
This spectrum revealed flare emission up to
an energy of $\sim$400 MeV.  The fundamental question is: what is the origin
of this emission?  The  nuclear line emission observed with the GBM implies
the presence of accelerated ions up to at least 50 MeV nucleon$^{-1}$.  It is
possible that the flare-accelerated proton spectrum extended up to the
$\sim$300 MeV threshold for pion production.  Alternatively, it is possible
that the LAT emission is from electron bremsstrahlung, either from an
extension to high energies of the electron spectrum producing the X-ray
bremsstrahlung observed in the GBM or from an additional hard electron
component.  One possible way to resolve this ambiguity is to jointly fit the
GBM and LAT spectra assuming different origins for the LAT emission.

\begin{figure}
\epsscale{.60}
\plotone{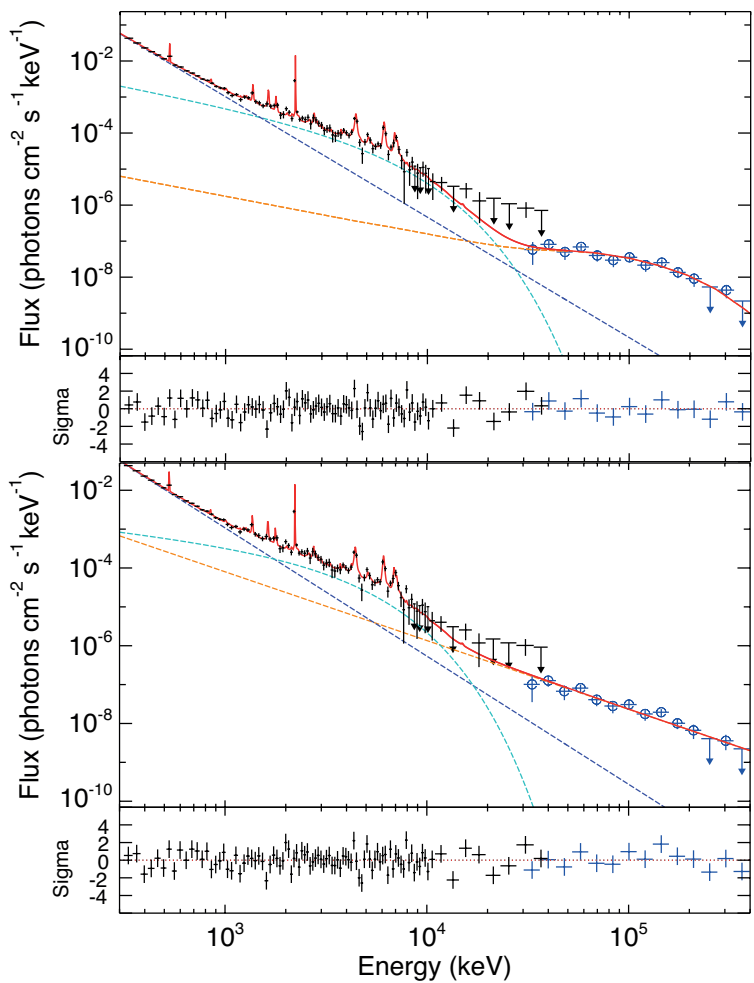}
\caption{
Combined GBM/LAT photon spectrum accumulated between 00:55:40 and 00:56:30
showing the best total fit using the same components as in Figure 3 plus an additional component for the LAT emission.  
The upper panel shows a pion-decay fit to the LAT spectrum; alternatively the lower panel shows a power-law
fit, presumedly representing a third electron bremsstrahlung component.
Note that because this is a photon representation the lines are plotted
at their intrinsic resolution and appear to be more significant than they really are.
}
\label{stack}
\end{figure}

In Figure \ref{stack} we plot the background-subtracted photon spectrum from
0.3 to 400 MeV including both the GBM and LAT data.  We made two fits to the
joint data sets, one  assuming that the observed LAT emission was from
pion-decay radiation (top panel) and the other assuming that it was from a
hard power-law spectrum of electron bremsstrahlung (bottom panel).   Based on
the statistical quality of the fits to the LAT spectrum we cannot distinguish
between the two emission models.  In addition we cannot constrain the origin
of the emission for this event by extrapolating the models into the GBM energy
range; however we note that for a stronger flare we might be able to rule out
a power-law model.  The choice of model also has only a small effect on the
parameters listed in Table 1 derived from fits to the 0.3 to 40 MeV GBM data.
Also plotted in the figure are the extensions into the LAT energy range of the
power-law and cutoff power-law components derived from the fits to the GBM
data.  The intensities of these components fall at least an order of magnitude
below the LAT measurements and therefore do not make a significant
contribution to the solar emission observed by the LAT.

%{\bf NOTE: The next two sentences are relocated from farther down in this section.}

Even though we cannot statistically distinguish between a pion-decay or electron-bremsstrahlung origin for the observed LAT emission, we can obtain the best-fitting parameters for  these components. If the LAT emission is from electron bremsstrahlung, we have shown that it cannot be a simple extension of the low-energy bremsstrahlung components that we determined from fits to the GBM data; it must be from a distinct population of electrons extending to energies of several hundred MeV. We have determined the power-law parameters (PL3) of the fitted bremsstrahlung spectrum and list
them in Table 1. However, this third high energy electron component cannot be produced by the acceleration models mentioned \S 4.1 above, which produce spectra that steepen beyond tens of MeV due to synchrotron energy losses that increase with energy
\citep[see][]{park97}, and must have a quite different origin.
Consequently we believe that this is a less likely scenario than the hadronic model.

Assuming that the LAT emission is from hadronic interactions, we have fit the LAT spectrum with calculated pion-decay spectra produced by accelerated ions having differential power-law indices from $-$2.5 to $-$7.5.  With 67\% confidence (based on $\chi^2$) we conclude that the spectrum of accelerated ions responsible for the pion-decay emission must be steeper than a power-law with index $-$4.5.  We note that there is no change in the quality of the fits for indices steeper than $-5$ due to limited statistics $>$400 MeV. We list the fluence of pion-decay photons $>$200 keV derived from these fits in Table 1.  This fluence is a factor of ten below the limit derived from fitting the GBM BGO data up to 40 MeV, illustrating the significantly greater sensitivity of the LAT instrument.

We can use the results of our GBM and LAT spectral analyses to obtain information on ions accelerated in the impulsive phase of the June 12 flare.  \citet{murp97} have described how parameters derived from integrated spectroscopic fits and temporal studies can be used to obtain this information.  We first use the nuclear de-excitation line, neutron-capture line, and pion-decay fluences listed in Table 1 to estimate the overall shape of the accelerated ion spectrum.  These three emissions are produced by accelerated ions within distinct energy ranges: $\sim$5-20 MeV for the de-excitation lines, $\sim$10-50 MeV for the neutron capture line, and $>$300 MeV for the pion-decay emission.  Ratios of these emissions therefore determine the relative numbers of accelerated ions in the associated energy ranges.  We then obtain spectral indices across these energy ranges by comparing measured ratios with ratios from theoretical calculations \citep{murp87,murp05,murp07} based on updated nuclear cross sections.

If we assume that the LAT emission $>$30 MeV was entirely due to pion-decay emission, then we estimate that the flare-accelerated ion spectrum was consistent with a series of power laws, softening with energy, with indices of $\sim$$-3.2$ between $\sim5-50$ MeV, $\sim$$-4.3$ between $\sim$50--300 MeV, and softer than $\sim$$-4.5$ above 300 MeV.  However, these calculations assume that these individual power laws continued without break to high energies, which cannot be the case where the spectrum softens with increasing energy.  We will describe a more refined representation of the accelerated proton spectrum in a forthcoming publication.

\subsection{Combined GBM and LAT Timing Studies}

The combined GBM and LAT observations also provide us with the opportunity to study acceleration and transport in this impulsive flare.  In Figure \ref{bnp2} we plot {\no 5-s resolution} time profiles of {\no the fitted} bremsstrahlung and nuclear de-excitation line fluxes from the GBM and the $>$100 MeV flux observed by the LAT.  For purposes of comparison we have plotted the bremsstrahlung profile over the nuclear and LAT $>$100 MeV histories in b) and c), respectively.  We have also plotted the nuclear profile over the LAT $>$100 MeV history. {\no The early peaking of the bremsstrahlung suggests that the higher-energy emissions were delayed by a few seconds and a cross-correlation study indicates that the overall lag between the LAT $>$100 MeV flux and the GBM 300 keV bremsstrahlung flux is $\sim$3 s.  

\begin{figure}
%\epsscale{.70}
\plotone{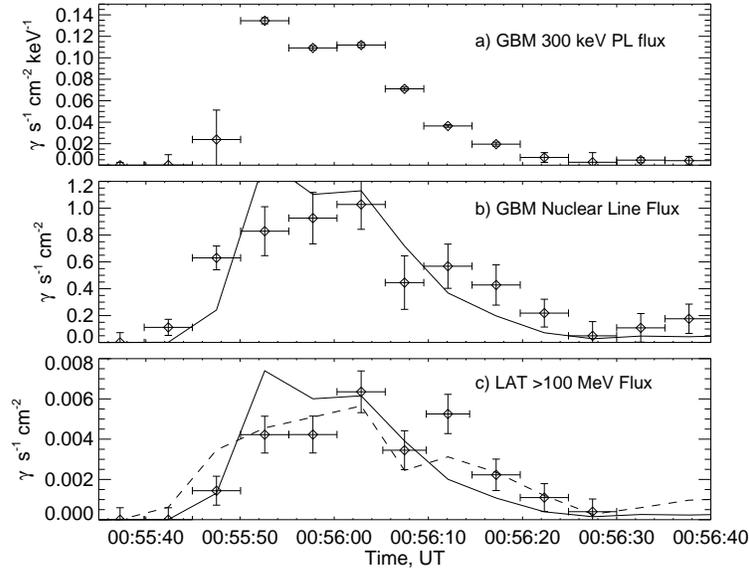}
\caption{Time profiles of the a) bremsstrahlung differential flux at 300 keV (GBM), b) total nuclear de-excitation line flux (GBM), and c) integral pion-decay flux $>$100 MeV (LAT). Solid curves over-plotted in b) and c) are the arbitrarily normalized bremsstrahlung flux plotted in a); dashed curve over-plotted in c) is arbitrarily normalized nuclear line flux from panel b). }
\label{bnp2}
\end{figure}

\begin{figure}
%\epsscale{.70}
\plotone{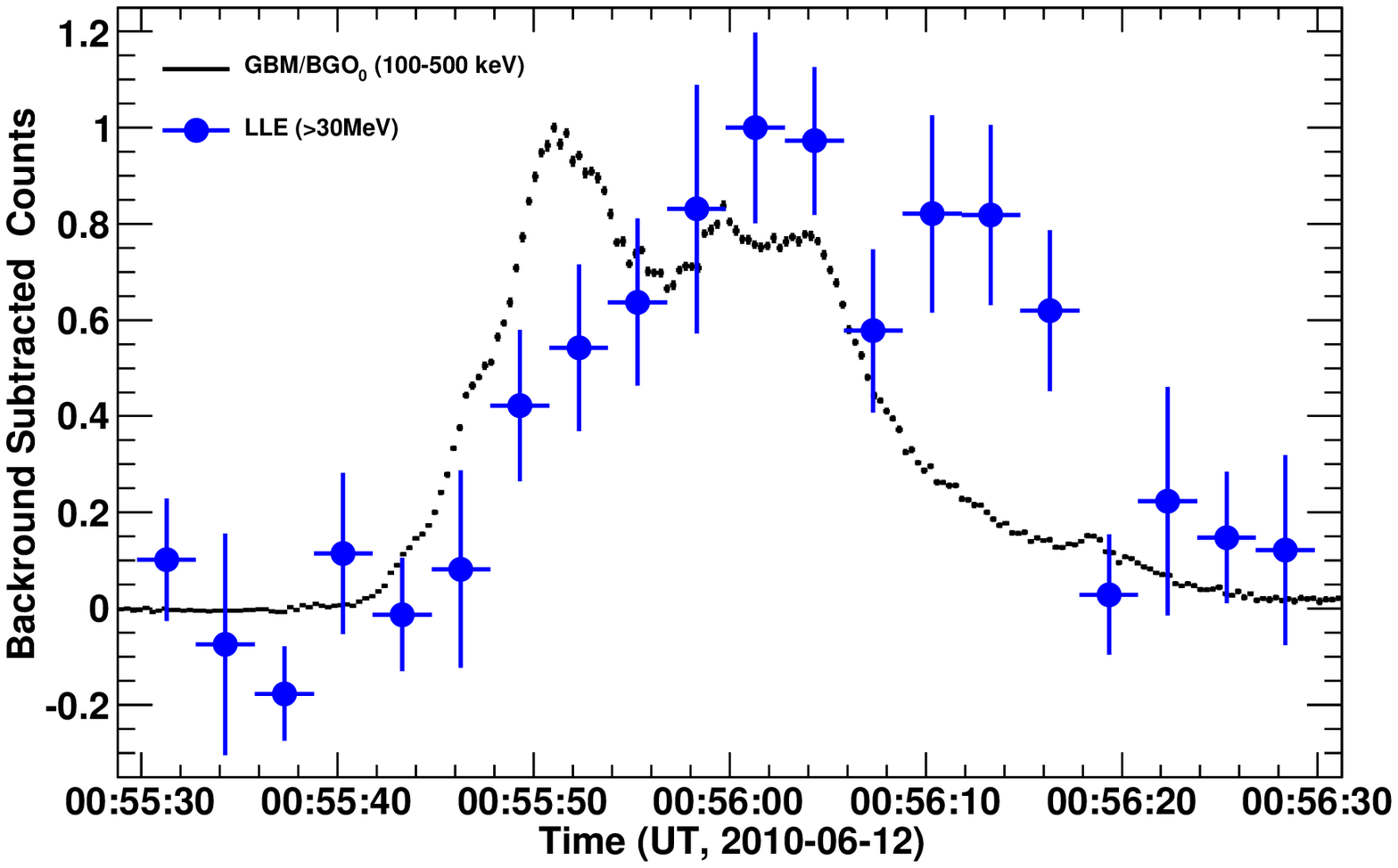}
\caption{\no{Comparison of high-time resolution profiles of the 100 -- 500 keV emission observed in the GBM BGO detector and of $>$30 MeV LLE data.  A cross-correlation analysis indicates that the high-energy $\gamma$-ray emission had an overall lag of 6 $\pm$ 3 s relative to the bremsstrahlung.} }
\label{xrlle}
\end{figure}

This delay warranted a higher-time resolution comparison of 100 -- 500 keV GBM and LAT $>$30 MeV counting rates rather than fitted fluxes.  The 100 -- 500 keV band is dominated by electron bremsstrahlung as can be seen in Figure \ref{gbm}.   In Figure \ref{xrlle} we plot GBM/BGO 100-500 keV rates at 320 ms resolution with 3-s LAT/LLE $>$30 MeV rates over plotted.  The hard X-ray profile reveals the presence of a clearly separated initial peak along with other structures.  The onset of the $>30$ MeV emission appears to be $\sim$3 s following the bremsstrahlung and rises to a peak about 10 s after the 100 --500 keV peak.  The LLE profile appears to reflect the double-peaked bremsstrahlung profile with a delay of about 10 seconds.  From a cross correlation analysis of the two profiles plotted in Figure \ref{xrlle} we find that the $>$30 MeV emission lags the bremsstrahlung by 6 $\pm$ 3 s.}
{\no There are two fundamental implications of the time profiles in Figure \ref{xrlle}:  1) protons and/or electrons began reaching energies above 100 MeV within a few seconds of the time it took to accelerate electrons to energies of hundreds of keV; and 2)  the overall acceleration time scale of the $>$100 MeV particles is similar to that observed in hundreds of keV electrons, but delayed by about 10 seconds.}

\subsection{Search for $>$100 MeV Emission Following the Flare}

The {\it{Fermi}} LAT detects quiescent emission from the Sun on a near daily basis \citep{abdo11}.  This emission comes from cosmic-ray proton interactions in the solar atmosphere and photosphere, and from Compton scattering of cosmic-ray electrons on solar blackbody photons.  The LAT is therefore a sensitive monitor of temporally extended solar-flare emission such as detected by {\it{CGRO}} EGRET experiment following the 1991 June 11 flare \citep{kanb93,rank01}.  We therefore studied the emission within 15$^{\circ}$ of the Sun in the hours preceding and following the flare.  The standard LAT data products were used in the analysis, which modeled the region around the Sun including all sources in the Second Fermi LAT Source Catalog~\citep{2FGL}, isotropic and Galactic diffuse emissions, and spatially extended Compton-scattered solar photons discussed above.

\begin{figure}
%\epsscale{.70}
\plotone{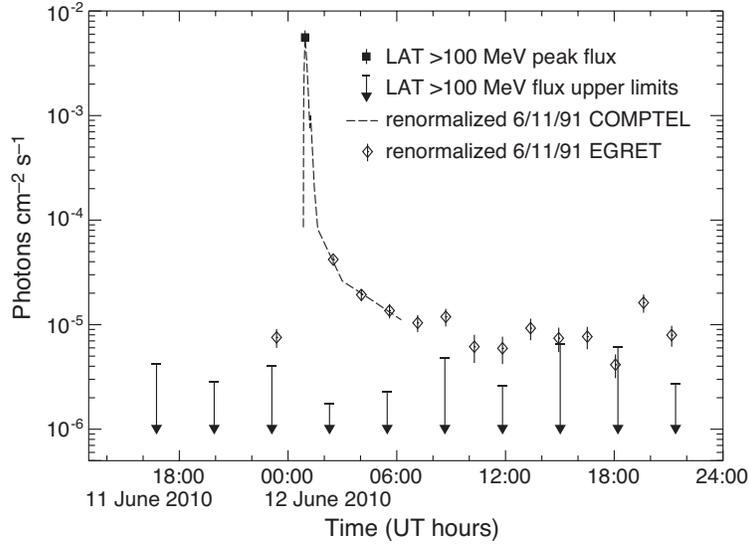}
\caption{95\% confidence upper limits of $>$100 MeV solar $\gamma$-rays measured by LAT within 7 hours preceding and 22 hours following the 2010 June 12 flare.  Dashed line is the 1991 June 11 time history of the 2.223 MeV neutron-capture line observed by COMPTEL experiment and diamonds are EGRET $>$50 MeV data both normalized to the peak $>$100 MeV flux measured by the LAT for the June 12 flare. The 2010 June 12 flare does not show extended emission like the 1991 June 11 flare observed with EGRET.}
\label{extended}
\end{figure}

In Figure~\ref{extended} we plot 95\% confidence limits on the $>$100 MeV flux from the solar disk  in $\sim$30 min exposures every two orbits from 6 hr before the flare to 22 hr after it. {\no  We note that the Sun was outside the FOV for LAT standard-product analysis during the flare, but its $\gamma$-ray emission could be studied in the orbits just before and after the flare.}
During this time period the Moon passed within 10$^{\circ}$ of the Sun and its flux of $\sim1 \times 10^{-6}~\gamma$ cm$^{-2}$ s$^{-1}$ made a significant contribution to the measured solar fluxes because we did not include it in the model of the source region.  There is no evidence for an increase in the  solar emission following the time of the flare, which is denoted by the peak flare flux $>$100 MeV.  This peak flux is about 1000 times higher than the plotted upper limits.  We wish to compare this to the time extended $>$100 MeV emission observed by EGRET following the 1991 June 11 flare \citep{kanb93,rank01}. But EGRET was saturated at the peak of the flare and therefore its time history cannot be normalized to the $>$100 MeV LAT peak flux of the June 12 flare.  However, after the peak of the flare the ratio of the $>$100 MeV EGRET flux to the 2.223 MeV neutron-capture line measured by COMPTEL, which operated normally throughout the June 11 flare, was relatively constant.  We therefore used the COMPTEL time history as a proxy for the EGRET $>$100 MeV photon fluxes.  The time profile of the COMPTEL 2.223 MeV neutron-capture line flux \citep{shar97}, normalized to the peak $>$100 MeV LAT June 12 flux, is shown in the Figure~\ref{extended}.
We see that during the first 30 minute exposure following the flare, the LAT upper limit is a factor of $\sim$ 20 below the value expected if there had been comparable extended emission similar to that found on 1991 June 11.

\section{Summary and Discussion}

The 2010 June 12 flare was the first in Cycle 24 to be observed to emit nuclear $\gamma$ rays.  It was also the first flare detected by the {\it{Fermi}} LAT at energies above 30 MeV.  The hard X-ray and nuclear line radiation was observed both by the {\it{Fermi}} GBM and {\it{RHESSI}} spectrometers.  In this paper we only analyzed GBM data because {\it{RHESSI}} was offset from the Sun to study the Crab Nebula during the time period of the flare; this affected knowledge of the instrument response.

The fact that the flare emitted detectable $\gamma$ rays at all is surprising because its peak soft X-ray emission only reached a {\it{GOES}} M2 level.  However, \citet{shih09} and others have shown that $\gamma$-ray line fluences are only weakly correlated with {\it{GOES}} soft X-ray emission but are strongly correlated with electron bremsstrahlung fluences $>$300 keV.  This is true for the June 12 flare as we find that the measured bremsstrahlung and 2.223 MeV fluences are consistent with the established correlation.

{\no The flare originated from a compact region and its hard X-ray emission only lasted 50 s.  Figure 6 reveals striking information about the processes that accelerate protons and/or electrons to energies of hundreds of MeV.  We find that although some of the particles reach energies $\gtrsim$100 MeV within about 3 s, the bulk of these particles reach such high energies following a delay of about 10 s.  This is revealed in the delayed double-peaked time structure $>$30 MeV that is similar to what is observed in hard X-rays.}  
 
 In Table 1 we list the best-fitting parameters from our fits {\no to the GBM and LAT spectra}.  These include the amplitudes (at 300 keV) and indices of two power-law continua observed by the GBM between 300 keV and 8 MeV.  The first is an extension of the hard X-ray spectrum observed by the GBM NaI detectors.  The second appears to be a hard power law with an exponential cutoff energy near 2.5 MeV. Although the GBM only has moderate spectral resolution, it was able to measure the fluences of the 0.511 MeV annihilation and 2.223 MeV neutron capture lines, and the total nuclear de-excitation emission.  There was no evidence in the GBM data for flare emission above about 8 MeV.  In contrast the LAT detected significant continuum emission from $\sim$30 to 400 MeV about an order of magnitude below the GBM upper limits. This reflects the excellent sensitivity of the LAT for observing solar flares.  This radiation could be either pion-decay emission or primary electron bremsstrahlung.  In Table 1 we list the measured pion-decay fluences $>$200 keV and $>$100 MeV.  We also list the parameters of a high-energy power-law bremsstrahlung component that fits the LAT data equally well.  Theoretical arguments made in \S 4.2 weigh against an electron origin for the emission observed by the LAT.  Under the assumption of an hadronic origin for the LAT emission and using $\gamma$-ray line measurements by the GBM, we have estimated the shape of the accelerated-ion spectrum that could have produced the combined spectrum.

We have also set significant constraints on any time-extended $>$100 MeV emission.  Our limit on the $>$100 MeV flux of photons is about an order of magnitude below what would have been expected if the decay followed that observed in the well-studied 1991 June 11 flare \citep{kanb93,rank01}.  Thus, there is no evidence for precipitation of trapped flare particles, particles accelerated in magnetic loops after the impulsive phase, particles accelerated in CME-associated reconnection sheets  \citep{ryan00}, or particles sharing the same origin as the Solar Energetic Particles (SEPs) observed in space \citep{ram87,cliv93}. We estimate from white light measurements of the 1991 June 11 flare \citep{saku92} that the foot-point separation was $\sim2.5 \times 10^4$ km, about 2.5 times larger than the June 12 flare.  It is possible that longer coronal loops are necessary for time-extended acceleration and/or trapping of protons.

This paper primarily addresses only the $\gamma$-ray observations of the 2010 June 12 flare.   Other studies are currently in progress involving hard X-ray observations of the flare by the {\it{Fermi}} GBM NaI detectors and comparisons of the characteristics of the ion and electron populations at the Sun and in space and at Earth.

\acknowledgments
{\no We wish to thank the referee for suggesting a more detailed examination of the delay between the hard X-ray bremsstrahlung and $>$30 MeV emission observed by LAT.}  The \textit{Fermi} LAT Collaboration acknowledges generous ongoing support
from a number of agencies and institutes that have supported both the
development and the operation of the LAT as well as scientific data analysis.
These include the National Aeronautics and Space Administration and the
Department of Energy in the United States, the Commissariat \`a l'Energie Atomique
and the Centre National de la Recherche Scientifique / Institut National de Physique
Nucl\'eaire et de Physique des Particules in France, the Agenzia Spaziale Italiana
and the Istituto Nazionale di Fisica Nucleare in Italy, the Ministry of Education,
Culture, Sports, Science and Technology (MEXT), High Energy Accelerator Research
Organization (KEK) and Japan Aerospace Exploration Agency (JAXA) in Japan, and
the K.~A.~Wallenberg Foundation, the Swedish Research Council and the
Swedish National Space Board in Sweden.

Additional support for science analysis during the operations phase is gratefully
acknowledged from the Istituto Nazionale di Astrofisica in Italy and the Centre National d'\'Etudes Spatiales in France.

Co-authors Briggs, Murphy, Schwartz, Share, and Tolbert were partially funded by the Fermi GI program to conduct the joint spectroscopic studies presented in this paper.


%% in parentheses, after the keyword, but they will not be verified.

%{\it Facilities:} \facility{Nickel}, \facility{HST (STIS)}, \facility{CXO (ASIS)}.



\bibliography{references}

\begin{thebibliography}{45}
\expandafter\ifx\csname natexlab\endcsname\relax\def\natexlab#1{#1}\fi

\bibitem[{{Abdo} {et~al.}(2010){Abdo}, {Ackermann}, {Ajello}, {Allafort},
  {Atwood}, {Baldini}, {Ballet}, {Barbiellini}, {Baring}, {Bastieri},
  {Baughman}, {Bechtol}, {Bellazzini}, {Berenji}, {Blandford}, {Bloom},
  {Bonamente}, {Borgland}, {Bouvier}, {Bregeon}, {Brez}, {Brigida}, {Bruel},
  {Burnett}, {Buson}, {Caliandro}, {Cameron}, {Caraveo}, {Carrigan},
  {Casandjian}, {Cecchi}, {{\c C}elik}, {Chekhtman}, {Cheung}, {Chiang},
  {Ciprini}, {Claus}, {Cohen-Tanugi}, {Conrad}, {Dermer}, {de Luca}, {de
  Palma}, {Dormody}, {Silva}, {Drell}, {Dubois}, {Dumora}, {Farnier},
  {Favuzzi}, {Fegan}, {Focke}, {Fortin}, {Frailis}, {Fukazawa}, {Funk},
  {Fusco}, {Gargano}, {Gasparrini}, {Gehrels}, {Germani}, {Giavitto},
  {Giebels}, {Giglietto}, {Giordano}, {Glanzman}, {Godfrey}, {Grenier},
  {Grondin}, {Grove}, {Guillemot}, {Guiriec}, {Hadasch}, {Harding}, {Hays},
  {Hobbs}, {Horan}, {Hughes}, {Jackson}, {J{\'o}hannesson}, {Johnson},
  {Johnson}, {Johnson}, {Kamae}, {Katagiri}, {Kataoka}, {Kawai}, {Kerr},
  {Kn{\"o}dlseder}, {Kuss}, {Lande}, {Latronico}, {Lee}, {Lemoine-Goumard},
  {Llena Garde}, {Longo}, {Loparco}, {Lott}, {Lovellette}, {Lubrano}, {Makeev},
  {Manchester}, {Marelli}, {Mazziotta}, {McConville}, {McEnery}, {McGlynn},
  {Meurer}, {Michelson}, {Mitthumsiri}, {Mizuno}, {Moiseev}, {Monte},
  {Monzani}, {Morselli}, {Moskalenko}, {Murgia}, {Nakamori}, {Nolan}, {Norris},
  {Noutsos}, {Nuss}, {Ohsugi}, {Omodei}, {Orlando}, {Ormes}, {Ozaki},
  {Paneque}, {Panetta}, {Parent}, {Pelassa}, {Pepe}, {Pesce-Rollins},
  {Pierbattista}, {Piron}, {Porter}, {Rain{\`o}}, {Rando}, {Ray}, {Razzano},
  {Reimer}, {Reimer}, {Reposeur}, {Ritz}, {Rochester}, {Rodriguez}, {Romani},
  {Roth}, {Ryde}, {Sadrozinski}, {Sander}, {Saz Parkinson}, {Sgr{\`o}},
  {Siskind}, {Smith}, {Smith}, {Spandre}, {Spinelli}, {Starck}, {Strickman},
  {Suson}, {Takahashi}, {Takahashi}, {Tanaka}, {Thayer}, {Thayer}, {Thompson},
  {Tibaldo}, {Torres}, {Tosti}, {Tramacere}, {Usher}, {Van Etten}, {Vasileiou},
  {Venter}, {Vilchez}, {Vitale}, {Waite}, {Wang}, {Watters}, {Weltevrede},
  {Winer}, {Wood}, {Ylinen}, \& {Ziegler}}]{vela2}
{Abdo}, A.~A., {et~al.} 2010, \apj, 713, 154

\bibitem[{{Abdo} {et~al.}(2011){Abdo}, {Ackermann}, {Ajello}, {Baldini},
  {Ballet}, {Barbiellini}, {Bastieri}, {Bechtol}, {Bellazzini}, {Berenji},
  {Bonamente}, {Borgland}, {Bouvier}, {Bregeon}, {Brez}, {Brigida}, {Bruel},
  {Buehler}, {Buson}, {Caliandro}, {Cameron}, {Caraveo}, {Casandjian},
  {Cecchi}, {Charles}, {Chekhtman}, {Chiang}, {Ciprini}, {Claus},
  {Cohen-Tanugi}, {Conrad}, {Cutini}, {de Angelis}, {de Palma}, {Dermer},
  {Digel}, {Silva}, {Drell}, {Dubois}, {Favuzzi}, {Fegan}, {Focke}, {Fortin},
  {Frailis}, {Funk}, {Fusco}, {Gargano}, {Gasparrini}, {Gehrels}, {Germani},
  {Giglietto}, {Giordano}, {Giroletti}, {Glanzman}, {Godfrey}, {Grenier},
  {Grillo}, {Guiriec}, {Hadasch}, {Hays}, {Hughes}, {Iafrate},
  {J{\'o}hannesson}, {Johnson}, {Johnson}, {Kamae}, {Katagiri}, {Kataoka},
  {Kn{\"o}dlseder}, {Kuss}, {Lande}, {Latronico}, {Lee}, {Lionetto}, {Longo},
  {Loparco}, {Lott}, {Lovellette}, {Lubrano}, {Makeev}, {Mazziotta}, {McEnery},
  {Mehault}, {Michelson}, {Mitthumsiri}, {Mizuno}, {Moiseev}, {Monte},
  {Monzani}, {Morselli}, {Moskalenko}, {Murgia}, {Nakamori}, {Naumann-Godo},
  {Nolan}, {Norris}, {Nuss}, {Ohsugi}, {Okumura}, {Omodei}, {Orlando}, {Ormes},
  {Ozaki}, {Paneque}, {Pelassa}, {Pesce-Rollins}, {Pierbattista}, {Piron},
  {Porter}, {Rain{\`o}}, {Rando}, {Razzano}, {Reimer}, {Reimer}, {Reposeur},
  {Ritz}, {Sadrozinski}, {Schalk}, {Sgr{\`o}}, {Share}, {Siskind}, {Smith},
  {Spandre}, {Spinelli}, {Strickman}, {Strong}, {Takahashi}, {Tanaka},
  {Thayer}, {Thayer}, {Thompson}, {Tibaldo}, {Torres}, {Tosti}, {Tramacere},
  {Troja}, {Uchiyama}, {Usher}, {Vandenbroucke}, {Vasileiou}, {Vianello},
  {Vilchez}, {Vitale}, {Vladimirov}, {Waite}, {Wang}, {Winer}, {Wood}, {Yang},
  \& {Ziegler}}]{abdo11}
---. 2011, \apj, 734, 116

\bibitem[{{Akimov} {et~al.}(1996){Akimov}, {Ambro{\v z}}, {Belov}, {Berlicki},
  {Chertok}, {Karlick{\'y}}, {Kurt}, {Leikov}, {Litvinenko}, {Magun},
  {Minko-Wasiluk}, {Rompolt}, \& {Somov}}]{akim96}
{Akimov}, V.~V., {et~al.} 1996, \solphys, 166, 107

\bibitem[{{Aschwanden}(2004)}]{asch04}
{Aschwanden}, M.~J. 2004, {Physics of the Solar Corona. An Introduction}, ed.
  {Aschwanden, M.~J.} (Praxis Publishing Ltd)

\bibitem[{{Atwood} {et~al.}(2009){Atwood}, {Abdo}, {Ackermann}, {Althouse},
  {Anderson}, {Axelsson}, {Baldini}, {Ballet}, {Band}, {Barbiellini}, \&
  et~al.}]{atwo09}
{Atwood}, W.~B., {et~al.} 2009, \apj, 697, 1071

\bibitem[{{Bissaldi} {et~al.}(2009){Bissaldi}, {von Kienlin}, {Lichti},
  {Steinle}, {Bhat}, {Briggs}, {Fishman}, {Hoover}, {Kippen}, {Krumrey},
  {Gerlach}, {Connaughton}, {Diehl}, {Greiner}, {van der Horst}, {Kouveliotou},
  {McBreen}, {Meegan}, {Paciesas}, {Preece}, \& {Wilson-Hodge}}]{biss09}
{Bissaldi}, E., {et~al.} 2009, Experimental Astronomy, 24, 47

\bibitem[{{Briggs} {et~al.}(2011){Briggs}, {Connaughton}, {Wilson-Hodge},
  {Preece}, {Fishman}, {Kippen}, {Bhat}, {Paciesas}, {Chaplin}, {Meegan}, {von
  Kienlin}, {Greiner}, {Dwyer}, \& {Smith}}]{brig11}
{Briggs}, M.~S., {et~al.} 2011, \grl, 38, L02808

\bibitem[{{Chupp} \& {Ryan}(2009)}]{chup09}
{Chupp}, E.~L., \& {Ryan}, J.~M. 2009, Research in Astronomy and Astrophysics,
  9, 11

\bibitem[{{Chupp} {et~al.}(1982){Chupp}, {Forrest}, {Ryan}, {Heslin}, {Reppin},
  {Pinkau}, {Kanbach}, {Rieger}, \& {Share}}]{chup82}
{Chupp}, E.~L., {et~al.} 1982, \apjl, 263, L95

\bibitem[{{Cliver} {et~al.}(1993){Cliver}, {Kahler}, \& {et al.}}]{cliv93}
{Cliver}, E.~W., {Kahler}, S.~W., \& {et al.} 1993, in International Cosmic Ray
  Conference, Vol.~3, International Cosmic Ray Conference, 91--+

\bibitem[{{Ellison} \& {Ramaty}(1985)}]{ell85}
{Ellison}, D.~C., \& {Ramaty}, R. 1985, \apj, 298, 400

\bibitem[{{Forrest} \& {Chupp}(1983)}]{forr83}
{Forrest}, D.~J., \& {Chupp}, E.~L. 1983, \nat, 305, 291

\bibitem[{{Forrest} {et~al.}(1986){Forrest}, {Vestrand}, {Chupp}, {Rieger}, \&
  {Cooper}}]{forr86}
{Forrest}, D.~J., {Vestrand}, W.~T., {Chupp}, E.~L., {Rieger}, E., \& {Cooper},
  J. 1986, Advances in Space Research, 6, 115

\bibitem[{{Forrest} {et~al.}(1985){Forrest}, {Vestrand}, {Chupp}, {Rieger},
  {Cooper}, \& {Share}}]{forr85}
{Forrest}, D.~J., {Vestrand}, W.~T., {Chupp}, E.~L., {Rieger}, E., {Cooper},
  J.~F., \& {Share}, G.~H. 1985, in International Cosmic Ray Conference,
  Vol.~4, International Cosmic Ray Conference, ed. {M.~Garcia-Munoz,
  K.~R.~Pyle, \& J.~A.~Simpson}, 146--149

\bibitem[{{Hoover} {et~al.}(2008){Hoover}, {Kippen}, {Wallace}, {Pendleton},
  {Fishman}, {Meegan}, {Kouveliotou}, {Wilson-Hodge}, {Bissaldi}, {Diehl},
  {Greiner}, {Lichti}, {von Kienlin}, {Steinle}, {Bhat}, {Briggs},
  {Connaughton}, {Paciesas}, \& {Preece}}]{hoov08}
{Hoover}, A.~S., {et~al.} 2008, in American Institute of Physics Conference
  Series, Vol. 1000, American Institute of Physics Conference Series, ed.
  {M.~Galassi, D.~Palmer, \& E.~Fenimore}, 565--568

\bibitem[{{Kanbach} {et~al.}(1993){Kanbach}, {Bertsch}, {Fichtel}, {Hartman},
  {Hunter}, {Kniffen}, {Kwok}, {Lin}, {Mattox}, \&
  {Mayer-Hasselwander}}]{kanb93}
{Kanbach}, G., {et~al.} 1993, \aaps, 97, 349

\bibitem[{{Kane} {et~al.}(1986){Kane}, {Chupp}, {Forrest}, {Share}, \&
  {Rieger}}]{kane86}
{Kane}, S.~R., {Chupp}, E.~L., {Forrest}, D.~J., {Share}, G.~H., \& {Rieger},
  E. 1986, \apjl, 300, L95

\bibitem[{{Kozlovsky} {et~al.}(1987){Kozlovsky}, {Lingenfelter}, \&
  {Ramaty}}]{koz87}
{Kozlovsky}, B., {Lingenfelter}, R.~E., \& {Ramaty}, R. 1987, \apj, 316, 801

\bibitem[{{Kozlovsky} {et~al.}(2004){Kozlovsky}, {Murphy}, \& {Share}}]{koz04}
{Kozlovsky}, B., {Murphy}, R.~J., \& {Share}, G.~H. 2004, \apj, 604, 892

\bibitem[{{Kuznetsov} {et~al.}(2011){Kuznetsov}, {Kurt}, {Yushkov}, {Kudela},
  \& {Galkin}}]{kuzn11}
{Kuznetsov}, S.~N., {Kurt}, V.~G., {Yushkov}, B.~Y., {Kudela}, K., \& {Galkin},
  V.~I. 2011, \solphys, 268, 175

\bibitem[{{Mart{\'{\i}}nez Oliveros} {et~al.}(2011){Mart{\'{\i}}nez Oliveros},
  {Couvidat}, {Schou}, {Krucker}, {Lindsey}, {Hudson}, \& {Scherrer}}]{oliv11}
{Mart{\'{\i}}nez Oliveros}, J.~C., {Couvidat}, S., {Schou}, J., {Krucker}, S.,
  {Lindsey}, C., {Hudson}, H.~S., \& {Scherrer}, P. 2011, \solphys, 269, 269

\bibitem[{{McTiernan} \& {Petrosian}(1990)}]{mcti90}
{McTiernan}, J.~M., \& {Petrosian}, V. 1990, \apj, 359, 541

\bibitem[{{Meegan} {et~al.}(2009){Meegan}, {Lichti}, {Bhat}, {Bissaldi},
  {Briggs}, {Connaughton}, {Diehl}, {Fishman}, {Greiner}, {Hoover}, {van der
  Horst}, {von Kienlin}, {Kippen}, {Kouveliotou}, {McBreen}, {Paciesas},
  {Preece}, {Steinle}, {Wallace}, {Wilson}, \& {Wilson-Hodge}}]{meeg09}
{Meegan}, C., {et~al.} 2009, \apj, 702, 791

\bibitem[{{Murphy} {et~al.}(1987){Murphy}, {Dermer}, \& {Ramaty}}]{murp87}
{Murphy}, R.~J., {Dermer}, C.~D., \& {Ramaty}, R. 1987, \apjs, 63, 721

\bibitem[{{Murphy} {et~al.}(2009){Murphy}, {Kozlovsky}, {Kiener}, \&
  {Share}}]{murp09}
{Murphy}, R.~J., {Kozlovsky}, B., {Kiener}, J., \& {Share}, G.~H. 2009, \apjs,
  183, 142

\bibitem[{{Murphy} {et~al.}(2007){Murphy}, {Kozlovsky}, {Share}, {Hua}, \&
  {Lingenfelter}}]{murp07}
{Murphy}, R.~J., {Kozlovsky}, B., {Share}, G.~H., {Hua}, X., \& {Lingenfelter},
  R.~E. 2007, \apjs, 168, 167

\bibitem[{{Murphy} {et~al.}(1997){Murphy}, {Share}, {Grove}, {Johnson},
  {Kinzer}, {Kurfess}, {Strickman}, \& {Jung}}]{murp97}
{Murphy}, R.~J., {Share}, G.~H., {Grove}, J.~E., {Johnson}, W.~N., {Kinzer},
  R.~L., {Kurfess}, J.~D., {Strickman}, M.~S., \& {Jung}, G.~V. 1997, \apj,
  490, 883

\bibitem[{{Murphy} {et~al.}(2005){Murphy}, {Share}, {Skibo}, \&
  {Kozlovsky}}]{murp05}
{Murphy}, R.~J., {Share}, G.~H., {Skibo}, J.~G., \& {Kozlovsky}, B. 2005,
  \apjs, 161, 495

\bibitem[{{Park} {et~al.}(1997){Park}, {Petrosian}, \& {Schwartz}}]{park97}
{Park}, B.~T., {Petrosian}, V., \& {Schwartz}, R.~A. 1997, \apj, 489, 358

\bibitem[{{Pelassa} {et~al.}(2010){Pelassa}, {Preece}, {Piron}, {Omodei},
  {Guiriec}, \& {Fermi LAT \& GBM collaborations}}]{pela10}
{Pelassa}, V., {Preece}, R., {Piron}, F., {Omodei}, N., {Guiriec}, S., \&
  {Fermi LAT \& GBM collaborations}. 2010, ArXiv e-prints

\bibitem[{{Petrosian} {et~al.}(1994){Petrosian}, {McTiernan}, \&
  {Marschhauser}}]{petr94}
{Petrosian}, V., {McTiernan}, J.~M., \& {Marschhauser}, H. 1994, \apj, 434, 747

\bibitem[{{Ramaty} {et~al.}(1987){Ramaty}, {Murphy}, \& {Dermer}}]{ram87}
{Ramaty}, R., {Murphy}, R.~J., \& {Dermer}, C.~D. 1987, \apjl, 316, L41

\bibitem[{{Rank} {et~al.}(2001){Rank}, {Ryan}, {Debrunner}, {McConnell}, \&
  {Sch{\"o}nfelder}}]{rank01}
{Rank}, G., {Ryan}, J., {Debrunner}, H., {McConnell}, M., \& {Sch{\"o}nfelder},
  V. 2001, \aap, 378, 1046

\bibitem[{{Reames}(1995)}]{ream95}
{Reames}, D.~V. 1995, Advances in Space Research, 15, 41

\bibitem[{{Rieger} \& {Marschh{\"a}user}(1991)}]{rieg91}
{Rieger}, E., \& {Marschh{\"a}user}, H. 1991, in Max '91/SMM Solar Flares:
  Observations and Theory, ed. {R.~M.~Winglee \& A.~L.~Kiplinger}, 68--+

\bibitem[{{Ryan}(2000)}]{ryan00}
{Ryan}, J.~M. 2000, \ssr, 93, 581

\bibitem[{{Sakurai} {et~al.}(1992){Sakurai}, {Ichimoto}, {Hiei}, {Irie},
  {Kumagai}, {Miyashita}, {Nishino}, {Yamaguchi}, {Fang}, {Kambry}, {Zhao}, \&
  {Shinoda}}]{saku92}
{Sakurai}, T., {et~al.} 1992, \pasj, 44, L7

\bibitem[{{Share} {et~al.}(1983){Share}, {Chupp}, {Forrest}, \&
  {Rieger}}]{shar83}
{Share}, G.~H., {Chupp}, E.~L., {Forrest}, D.~J., \& {Rieger}, E. 1983, in
  American Institute of Physics Conference Series, Vol. 101, Positron-Electron
  Pairs in Astrophysics, ed. {M.~L.~Burns, A.~K.~Harding, \& R.~Ramaty}, 15--20

\bibitem[{{Share} {et~al.}(1997){Share}, {Murphy}, \& {Ryan}}]{shar97}
{Share}, G.~H., {Murphy}, R.~J., \& {Ryan}, J. 1997, in American Institute of
  Physics Conference Series, Vol. 410, Proceedings of the Fourth Compton
  Symposium, ed. {C.~D.~Dermer, M.~S.~Strickman, \& J.~D.~Kurfess}, 17--36

\bibitem[{{Shih} {et~al.}(2009){Shih}, {Lin}, \& {Smith}}]{shih09}
{Shih}, A.~Y., {Lin}, R.~P., \& {Smith}, D.~M. 2009, \apjl, 698, L152

\bibitem[{{The Fermi-LAT Collaboration}(2011)}]{2FGL}
{The Fermi-LAT Collaboration}. 2011, ArXiv e-prints

\bibitem[{{Thompson} {et~al.}(1993){Thompson}, {Bertsch}, {Fichtel}, {Hartman},
  {Hofstadter}, {Hughes}, {Hunter}, {Hughlock}, {Kanbach}, {Kniffen}, {Lin},
  {Mattox}, {Mayer-Hasselwander}, {von Montigny}, {Nolan}, {Nel}, {Pinkau},
  {Rothermel}, {Schneid}, {Sommer}, {Sreekumar}, {Tieger}, \&
  {Walker}}]{Thompson1993}
{Thompson}, D.~J., {et~al.} 1993, \apjs, 86, 629

\bibitem[{{Vestrand} {et~al.}(1999){Vestrand}, {Share}, {Murphy}, {Forrest},
  {Rieger}, {Chupp}, \& {Kanbach}}]{vest99}
{Vestrand}, W.~T., {Share}, G.~H., {Murphy}, R.~J., {Forrest}, D.~J., {Rieger},
  E., {Chupp}, E.~L., \& {Kanbach}, G. 1999, \apjs, 120, 409

\bibitem[{{Vilmer} {et~al.}(2003){Vilmer}, {MacKinnon}, {Trottet}, \&
  {Barat}}]{vilm03}
{Vilmer}, N., {MacKinnon}, A.~L., {Trottet}, G., \& {Barat}, C. 2003, \aap,
  412, 865

\bibitem[{{Young} {et~al.}(2001){Young}, {Arndt}, {Bennett}, {Connors},
  {Debrunner}, {Diehl}, {McConnell}, {Miller}, {Rank}, {Ryan}, {Schoenfelder},
  \& {Winkler}}]{young01}
{Young}, C.~A., {et~al.} 2001, in American Institute of Physics Conference
  Series, Vol. 587, Gamma 2001: Gamma-Ray Astrophysics, ed. {S.~Ritz,
  N.~Gehrels, \& C.~R.~Shrader}, 613--617

\end{thebibliography}

\end{document}